\documentclass[twocolumn,twoside]{revtex4-1}
\usepackage{graphicx}
\usepackage{fancyhdr}
\usepackage{ amssymb }
\begin{document}

\title{Rare strange particle decays}

%

\author{M. Zamkovsky \\ on behalf of the~NA62 Collaboration \footnote{
R.~Aliberti, F.~Ambrosino, R.~Ammendola, B.~Angelucci, A.~Antonelli, G.~Anzivino, R.~Arcidiacono,
M.~Barbanera, A.~Biagioni
% ...
}}
\affiliation{Charles University, Prague, Czech Republic}

\begin{abstract}
The~rare decays $\mathrm{K}^+\to \pi^+ \nu \bar{\nu}$ and $\mathrm{K_L}\to \pi^0 \nu \bar{\nu}$ are extremely attractive processes to study
flavor physics because they are both exceptionally clean from a~theoretical point of view
These modes are measured by the~experiments NA62 at CERN in Switzerland and KOTO at J-PARC in Japan, respectively.
The~latest results from these experiments together with future prospects are presented.
The~NA62 experiment has, besides the~main goal, a~rich physics program on other rare kaon decays which will be also discussed.
\end{abstract}

\maketitle

\thispagestyle{fancy}


\section{Introduction}\label{Intro}
Kaon physics is still in the~main scope of both theoretical and experimental physics.
The~ultra~rare kaon decays $\mathrm{K}^+\to \pi^+ \nu \bar{\nu}$ and $\mathrm{K_L}\to \pi^0 \nu \bar{\nu}$ are particularly of interest.
They have very precise theoretical prediction that makes them sensitive to physics beyond the~Standard Model (SM).
So far only the~charged mode has been observed by the~experiments E787 and E949 at Brookhaven National Laboratory \cite{Artamonov:2009sz}.
The~result obtained with a~kaon decay-at-rest technique is: $BR(\mathrm{K^+}\to\pi^+\nu\bar{\nu}) = (17.3^{+11.5}_{-10.5}) \times 10^{-11} $, consistent with the~SM expectation within its large statistical uncertainty.
The~neutral mode has an upper limit at 90\% CL $BR(\mathrm{K_L}\to\pi^0\nu\bar{\nu}) < 2.6 \times 10^{-8}$ set by the~E391a~experiment \cite{E391}.
The~NA62 experiment aims to measure the~branching ratio of the~$\mathrm{K}^+\to \pi^+ \nu \bar{\nu}$ decay with an error comparable to the~theoretical prediction
and the~KOTO experiment wants to make an observation of the~neutral mode.

The~NA62 experimental apparatus can measure a~variety of other processes besides the~main goal.
Specifically, the~further physical program includes search for Heavy Neutral Leptons (HNL), Lepton Flavour and Lepton Number Violation (LFV and LNV), searches for dark photons, inflatons, etc.

\section{Physics Motivation}\label{Physiscs}
The~$\mathrm{K}^+\to \pi^+ \nu \bar{\nu}$ and $\mathrm{K_L}\to \pi^0 \nu \bar{\nu}$ modes are flavor changing neutral current (FCNC) decays that are theoretically exceptionally clean.
They proceed through box and electroweak penguin diagrams and, thanks to a~quadratic GIM mechanism and a~strong Cabibbo suppression, are extremely rare.
Using the~value of tree-level elements of  the~Cabibbo-Kobayashi-Maskawa (CKM)  matrix  as  external  inputs,  the~SM predicts \cite{Brod, Buras:2015qea}:
\begin{equation}
  \begin{array}{lll}
    & BR(\mathrm{K^+}\to\pi^+\nu\bar{\nu})  &= (8.4 \pm 1.0) \cdot 10^{-11},
   \\
    & BR(\mathrm{K_L}\to\pi^0\nu\bar{\nu}) &= (3.4 \pm 0.6) \cdot 10^{-11}.
  \end{array}
\end{equation}

The~theoretical origin of these processes is dominated by short distance contribution without hadronic uncertainties as the~hadronic matrix elements are extracted from the~well-known decay $\mathrm{K^+}\to \pi^0 e^+\nu$.
As the~perturbative QCD and electroweak corrections in both decays are under full control, the~theoretical error budget within the~SM is dominated by the~CKM parameter uncertainties.

The~$\mathrm{K}\to \pi \nu \bar{\nu}$ decays are extremely sensitive to physics beyond the~SM, probing the~highest mass scales among the~rare meson decays.
Large deviations are expected in several beyond the~SM scenarios \cite{Blanke:2015wba}.
For example, the~tree-level FCNC mediated by heavy gauge boson ($Z'$) \cite{Buras:2015yca}, should show a~specific correlations between  $BR(\mathrm{K}^+\to \pi^+ \nu \bar{\nu})$ and $BR(\mathrm{K_L}\to \pi^0 \nu \bar{\nu})$.
In the~recent models with Lepton Flavour Universality violation \cite{Bordone:2017lsy}, the~anomalies observed in the~semileptonic B meson decays could be linked to $BR(\mathrm{K}^+\to \pi^+ \nu \bar{\nu})$ measurement.

\section{Experimental Setup}

\subsection{NA62}
\begin{figure}[h]
 \includegraphics[width=.5\textwidth]{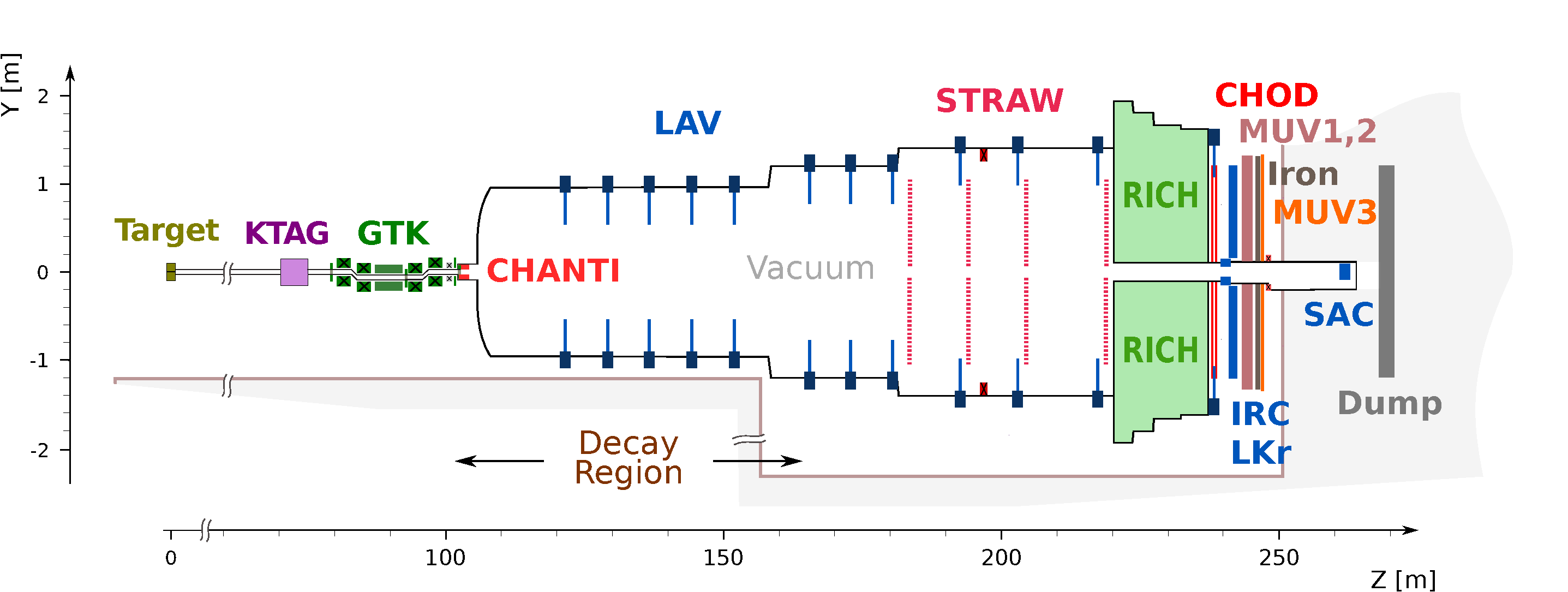}%
 \caption{Schematic side view of the~NA62 experiment. \label{NA62layout}}
\end{figure}

The~NA62 experiment uses the~protons from the~CERN SPS accelerator, which impinge on a~beryllium target and produce a~secondary hadron beam at nominal momentum of 75~GeV/$c$ and 1\% momentum spread (rms) at 750~MHz, about 6\% of which are kaons. 
The~beam is accompanied by a~muon halo at the~nominal rate of~3~MHz in the~detector acceptance.
The~first upstream detectors are the~Differential Cerenkov counter (KTAG) used to identify $\mathrm{K^+}$ in the~beam, the~silicon pixel beam spectrometer GTK and the~Charged Anti-counter detector, 
used to suppress products of inelastic interactions in the~GTK.
They are followed by a~110~m long vacuum tank dfining the~fiducial volume (FV) in first 60~m, where about~13\% of the~$\mathrm{K^+}$ entering the~experiment decay.
It contains the~four tracking stations of~the~magnetic spectrometer. 
Around the~vacuum tube twelve ring-shaped Large Angle Vetoes (LAV) are located with increasing diameter with distance from the~target, which together with the~electromagnetic Liquid Krypton calorimeter (LKr),
the~Inner Ring Calorimeter (IRC) and Small Angle Calorimeter (SAC) provide hermetic acceptance for photons emitted in the~$\mathrm{K}^+$ decays for polar angle up to 50 mrad.
A~Ring-Imaging Cherenkov counter (RICH) for~particle identification and two plastic scintillator charged hodoscopes (CHOD) are located downstream of the~vacuum tank, before the~LKr.
For further particle identification two hadronic calorimeters and muon veto detector (MUV1,2,3) follow the~LKr.
Additional counters (MUV0, HASC) installed at optimized locations provide hermetic coverage for charged particles produced in multi-track kaon decays.
The~schematic layout of the~detector is shown in Fig. \ref{NA62layout} and the~full detector description can be found in \cite{NA62:2017rwk}.

\subsection{KOTO}
\begin{figure}[h]
 \includegraphics[width=.5\textwidth]{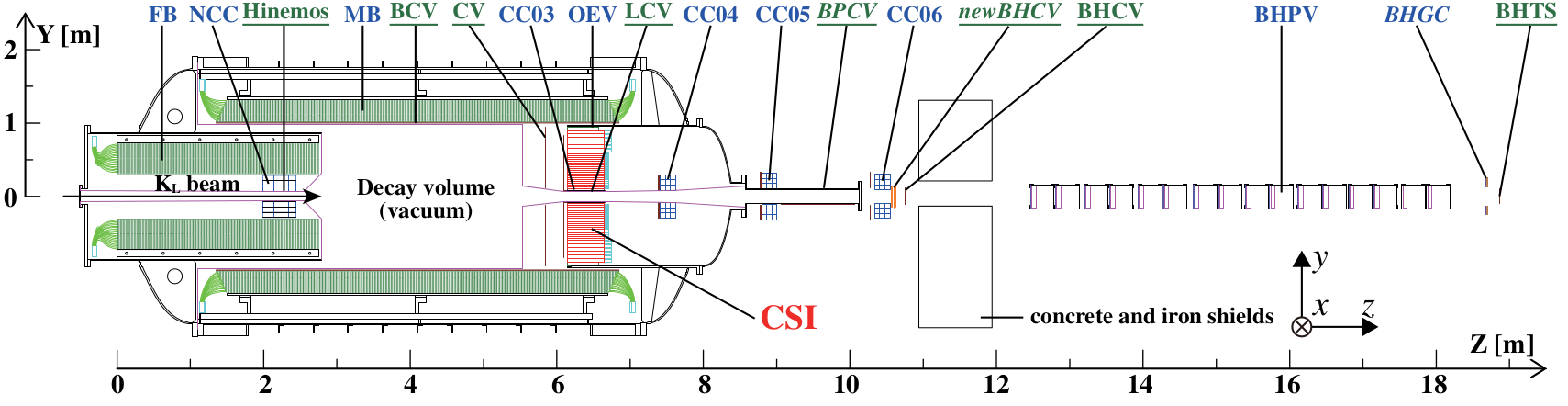}%
 \caption{Layout of the~KOTO experiment. Detector components with their abbreviated names written in blue (in green and underlined) are photon (charged particle) veto counters. \label{KOTOlayout}}
\end{figure}

The~KOTO experiment [12] at J-PARC is a~fixed target detector using primary protons at 30~GeV/$c$ extracted from the~J-PARC Main Ring accelerator and directed onto a~66~mm-long gold target.
The~secondary $\mathrm{K_L}$ beam, accompanied with neutrons and photons, with a~peak momentum of 1.4~GeV/$c$ is produced at an angle of 16~degrees with respect to the~nominal beam.
A~halo of neutrons produced in the~collimators, traveling outside of the~nominal beam solid angle, are present.
The~decay volume is a~3~m long chamber evacuated down to $5\times 10^{-7}$~mbar to suppress $\pi^0$ produced in the~interactions of beam neutrons with the~residual gas.
The~signature of $\mathrm{K_L}\to \pi^0 \nu \bar{\nu}$ are two photons from $\pi^0$ and nothing else, so the~main signal detector is an electromagnetic calorimeter CSI.
It consist of 2716 undoped CsI crystals, 50~cm long and $2.5\times 2.5 (5 \times 5)$  $\mathrm{cm^2}$ cross section within (outside) the~central $1.2 \times 1.2$~$\mathrm{m^2}$ region, 
with a~hole in the~middle to let the~beam particles pass through.
The~veto counters outside of the~decay region consist of lead-scintillator sandwich, lead-aerogel, or undoped-CsI counters for photons and plastic scintillators or wire chambers for charged particles.
After the~2013 Physics run, KOTO implemented several improvements to the~experimental apparatus.
These include: reduced thickness of the~vacuum window from 125~$\mu$m to 12~$\mu$m; added beam profile monitor for better beam alignment; added beam pipe charged veto.
The~detailed description of the~experimental setup can be found in \cite{Yamanaka:2012yma, Ahn:2016kja} and the~schematic layout is in Fig. \ref{KOTOlayout}.

\section{The~NA62 $\mathrm{K}^+\to \pi^+ \nu \bar{\nu}$ analysis}
The~signature of the~signal is one charged track in GTK compatible with a~kaon hypothesis and one in the~detector downstream.  
Kaon decays and beam−related activity are sources of background.
During the~2016 data taking, a~sample of $N_K = 1.1(2) \times 10^{11}$ was collected and further analyzed.

\subsection{Event Selection}
\begin{figure}[h]
 \includegraphics[width=.5\textwidth]{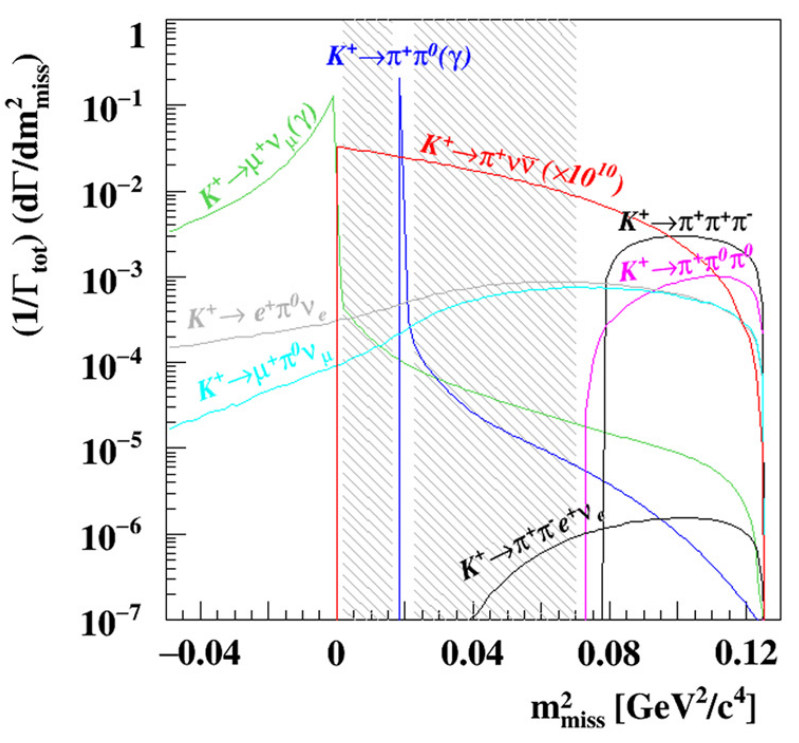}%
 \caption{ True $m_{miss}^2$ distribution of the~$\mathrm{K}^+\to \pi^+ \nu \bar{\nu}$ decay together with the~main $\mathrm{K^+}$ decay modes, computed under $\pi^+$ mass hypothesis in the~final state.
           The~signal (red) is multiplied by a~factor of $10^{10}$ for a~better visibility, and dashed areas show the~signal regions.\label{Figure4}}
\end{figure}

\begin{figure}[h]
 \includegraphics[width=.5\textwidth]{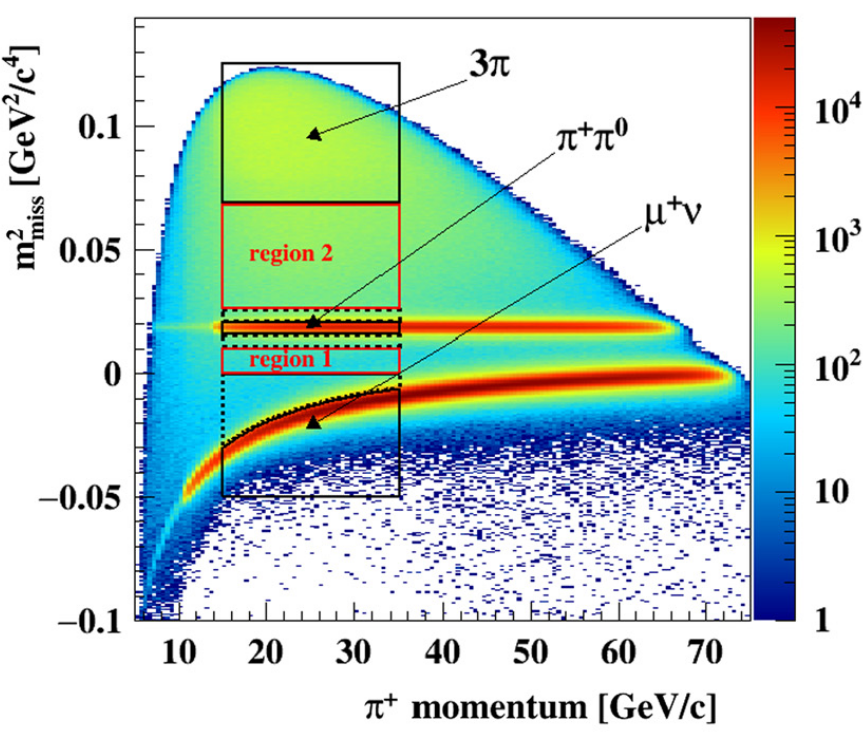}%
 \caption{Distribution of the~$m_{miss}^2$ as a~function of the~$\pi^+$ momentum based on a~control data sample. The~main backgrounds together with the~signal regions are highlighted. \label{Figure5}}
\end{figure}

The~squared missing mass calculated from an upstream $\mathrm{K^+}$ 4-momentum $p_K$ and a~downstream $\pi^+$ 4-momentum $p_{\pi}$ ($m_{miss}^2 = (p_K-p_{\pi})^2$) is used to discriminate kinematically the~main $\mathrm{K^+}$ 
decay modes from the~signal.
Based on $m_{miss}^2$ spectrum, two signal regions on either side of the~$\mathrm{K^+}\to \pi^+ \pi^0$ peak are defined (see Fig. \ref{Figure4}).
The~signal events must have a~$\pi^+$ momentum in the~(15,35) GeV/$c$ range.
The~distribution of $m_{miss}^2$ as a~function of the~$\pi^+$ momentum from a~control data sample is depicted in Fig. \ref{Figure5}.
Kinematic selection is not powerful enough to separate signal from background events, so the~additional suppression by $\pi^+$ identification (ID), photon and multiplicity rejection is required.
The~particle identification is based on the~RICH measurement and a~multivariate analysis of calorimetric information from LKr, MUV1 and MUV2, with MUV3 information used as veto.
The~$\pi^+$ ID efficiency within (15,35) GeV/$c$ momentum range is 78\% (82\%) using calorimeters (RICH), and the~corresponding muon mis-ID as $\pi^+$ is $0.6 \times 10^{-5}\ (2.1 \times 10^{-3})$, respectively.

Photon rejection requires no in-time activity in any of the~LAV, IRC, SAC and no additional clusters in the~LKr beyond 100~mm radius around the~$\pi^+$ impact point.
Multiplicity rejection criteria against photons interacting in the~detector material upstream of the~LKr include:
no in-time activity in the~hodoscopes unrelated to the~$\pi^+$ but in spatial coincidence with an energy deposit of at least 40~MeV in the~LKr;
no additional segments reconstructed in the~STRAW compatible with the~decay vertex;
no in-time signals in HASC and MUV0;
fewer than four extra signals in the~hodoscope in time with the~$\pi^+$.
The~resulting $\pi^0 \to \gamma \gamma$ rejection inefficiency is measured to be $2.5 \times 10^{-8}$ at 10\% (24\%) signal loss due to $\pi^+$ interactions (accidentals).
Photon and multiplicity rejection is also effective against $\mathrm{K^+}\to \pi^+ \pi^+ \pi^-$ and $\mathrm{K^+}\to \pi^+ \pi^- e^+ \nu$ backgrounds.

\subsection{Single Event Sensitivity}
The~single event sensitivity is defined as $SES = 1/(N_K \cdot \varepsilon_{\pi \nu \nu})$, where $N_K$ is the~number of $\mathrm{K^+}$ decays in the~fiducial volume and $\varepsilon_{\pi \nu \nu}$ is the~signal efficiency,
taking into account the~selection acceptance, trigger efficiency and photon plus multiplicity rejection inefficiency.
The~particle identification inefficiency is included in the~selection acceptance $A_{\pi\nu\nu} = (4.0 \pm 0.1) \%$, which is evaluated using MC simulation.
The~number of kaons was obtained from the~normalization mode $\mathrm{K^+}\to \pi^+ \pi^0$ selected in the~same way as the~signal except for photon and multiplicity rejection and requiring $m_{miss}^2$ to be in the~$\pi^+ \pi^0$ region.

The~SES and the~corresponding number of SM $\mathrm{K}^+\to \pi^+ \nu \bar{\nu}$ decays expected in the~signal regions are:
\begin{eqnarray}
 &SES       &= (3.15 \pm 0.01_{stat}  \pm 0.24_{syst}) \times 10^{-10},   \\
 &N^{exp}_{\pi\nu\nu}(SM)     &= 0.267\pm 0.001_{stat} \pm 0.020_{syst} \pm 0.032_{ext}. \nonumber
\label{eq-sp}
\end{eqnarray}
The~systematic uncertainty includes those of $A_{\pi\nu\nu}$, trigger efficiency, photon and multiplicity rejection inefficiency.
The~external error refers to uncertainties of the~SM parameters.

\subsection{Expected Background and Results}
Background from $\mathrm{K^+}$ decays in the~FV is mainly due to $\mathrm{K^+}\to \pi^+ \pi^0 (\gamma)$, $\mathrm{K^+}\to \mu^+ \nu (\gamma)$, $\mathrm{K^+}\to \pi^+ \pi^+ \pi^-$ and $\mathrm{K^+}\to \pi^+ \pi^- e^+ \nu$ decays.
Other kaon decays give negligible contribution, estimated at $\mathcal{O}(10^{-3})$. 
The~upstream backgrounds are due to pions originating upstream of the~fiducial volume.
They are produced in: $\mathrm{K^+}$ decays between GTK stations 2 and 3, matching an accidental beam particle; elastic scattering of beam $\pi^+$ in GTK 2 and 3, matched to an accidental $\mathrm{K^+}$;
$\mathrm{K^+}$ interactions with beam line material, produced either promptly or as a~decay product of a~neutral kaon.

The~table \ref{bkg} gives a~summary of all background sources.
\begin{table}[h]
\begin{center}
\caption{Summary of the~background estimates summed over the~two signal regions.}
   \begin{tabular}{l|l}
  \hline
   Process                                          &  Expected events in R1+R2 \\
   \hline
     $\mathrm{K^+}\to\pi^+\pi^0(\gamma)$ IB          &  $0.064 \pm 0.007_{stat} \pm0.006_{syst}$  \\
     $\mathrm{K^+}\to\mu^+\nu(\gamma)$ IB            &  $0.020 \pm 0.003_{stat} \pm0.006_{syst}$      \\
     $\mathrm{K^+}\to\pi^+\pi^-e^+\nu$               &  $0.013^{+0.017}_{-0.012}|_{stat} \pm0.009_{syst}$  \\
     $\mathrm{K^+}\to\pi^+\pi^+\pi^-$                &  $0.002 \pm 0.001_{stat} \pm0.002_{syst}$  \\
     $\mathrm{K^+}\to\pi^0\mu^+\nu$, $\mathrm{K^+}\to\pi^0\mu^+\nu$ & $<0.001$ \\
     $\mathrm{K^+}\to\pi^+\gamma\gamma$              &  $< 0.002$ \\
     Upstream background                             &  $0.050^{+0.090}_{-0.030}|_{stat}$ \\
     \hline
     Total Background                                &  $0.152^{+0.092}_{-0.033}|_{stat} \pm0.013_{syst}$      \\
     \hline
  \end{tabular}
  \label{bkg}
\end{center}
\end{table}

\begin{figure}[h]
 \includegraphics[width=.5\textwidth]{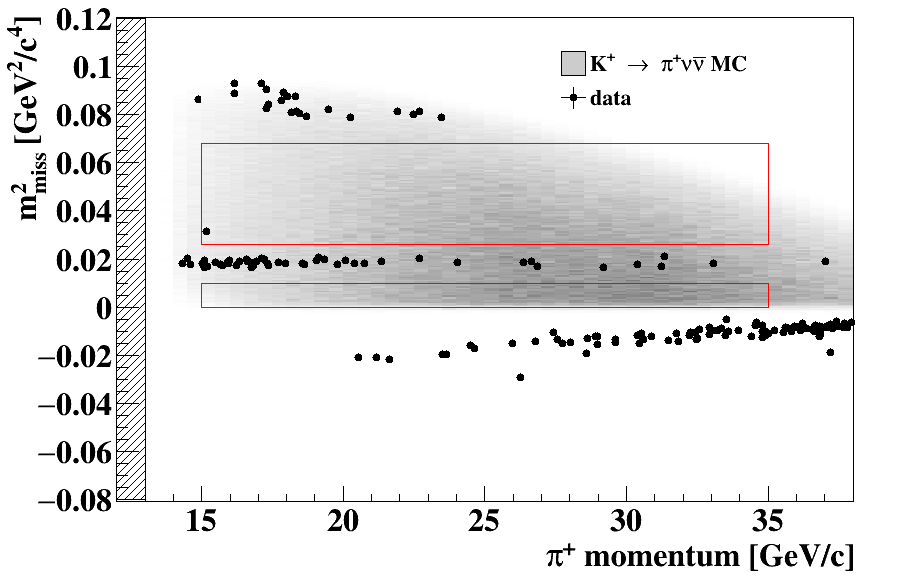}%
 \caption{Reconstructed  $m_{miss}^2$ as a~function of momentum of $\pi^+$ (markers) satisfying the~$\mathrm{K^+}\to \pi^+ \nu \bar{\nu}$ selection, except the~$m_{miss}^2$ and $\pi^+$ momentum criteria.
The~grey area corresponds to the~expected distribution of $\mathrm{K}^+\to \pi^+ \nu \bar{\nu}$ MC events. Red contours define the~signal regions. \label{signal}}
\end{figure}

After unmasking the~signal regions, one event is found in Region 2, as shown in Fig. \ref{signal}.

Considering the~observation of one event, the~$p$-value of the~signal and background hypothesis is 15\% and the~corresponding observed upper limit is:

\begin{equation}
 BR(\mathrm{K^+}\to\pi^+\nu\bar{\nu}) < 14 \times 10^{-10}\ \mathrm{at}\ 95\%\ \mathrm{CL}.
\end{equation}

A~detailed description of the~analysis procedure and background estimation can be found in \cite{CortinaGil:2018fkc}.

\subsection{Future Prospects}
The~2016 result is based on 2\% of the~total NA62 exposure in 2016–2018 and demonstrates the~validity of the~decay-in-flight technique in terms of background rejection and in view of the~measurement in progress using the~full data sample.
The~analysis of 2017 data sample is ongoing with the~2016-like event selection, hence has a~comparable analysis performance.
However, there are some improvements in the~treatment of the~pileup in IRC and SAC resulting in 40\% lower $\pi^0$ rejection inefficiency and slightly improved usage of RICH variables.
The~2017 data set also allows the~detailed comparison between signal and background shapes in the~signal regions as a~function of $\pi^+$ momentum.
The~preliminary estimations are summarized in table \ref{2017}.

\begin{table}

 \caption{Preliminary estimation of expected SM $\mathrm{K^+}\to\pi^+\nu\bar{\nu}$ events and background contributions based on 2017 data sample.}
\begin{center}
   \begin{tabular}{l|l}
      \multicolumn{2}{c}{ PRELIMINARY} \\
  \hline
   $N_K$                                                    &  $(13\pm1)\times 10^{11}$ \\
   \hline
   $SES$                                                    &  $(0.34\pm0.04) \times 10^{-10}$ \\
   \hline
    $\mathrm{K^+}\to\pi^+\nu\bar{\nu}$ &  $2.5  \pm 0.4$      \\
    \hline
    \hline 
      \multicolumn{2}{c}{ Background} \\
      \hline
     $\mathrm{K^+}\to\pi^+\pi^0(\gamma)$ IB          &  $0.35  \pm 0.02_{stat} \pm0.03_{syst}$  \\
     $\mathrm{K^+}\to\mu^+\nu(\gamma)$ IB            &  $0.16  \pm 0.01_{stat} \pm0.05_{syst}$      \\
     $\mathrm{K^+}\to\pi^+\pi^-e^+\nu$               &  $0.22  \pm 0.08_{stat} $  \\
     $\mathrm{K^+}\to\pi^+\pi^+\pi^-$                &  $0.015 \pm 0.008_{stat} \pm0.015_{syst}$  \\
     $\mathrm{K^+}\to\pi^+\gamma\gamma$              &  $0.005 \pm 0.005_{syst}$  \\
     $\mathrm{K^+}\to\ell^+\pi^0\nu_{\ell}$          &  $0.012 \pm 0.012_{syst}$  \\
     Upstream background                             &  Analysis on-going \\
     \hline
  \end{tabular}
 \end{center} 
 \label{2017}
\end{table}

\section{The~KOTO $\mathrm{K_L}\to \pi^0 \nu \bar{\nu}$ analysis}
The~KOTO experiment collected $N(K_L) \sim 2.4 \times 10^{11}$ and reached single event sensitivity at the~level of $(SES 1.3\times 10^{-8})$ in 2013.
An upper limit on BR$(\mathrm{K_L}\to\pi^0\nu\bar{\nu})$ was set to $<5.1\times 10^{-8}$ at 90\% CL \cite{Ahn:2016kja}.

During 2015 a~data set of $2.2 \times 10^{19}$ protons on target was collected.
The~event selection is based on the~$\pi^0$ reconstruction of from two clusters above 3~MeV in CSI.
The~$\pi^0$ decay vertex $Z_{vtx}$ and transverse momentum $P_{t}$ were calculated assuming that the~vertex was on the~beam axis.
The~$Z_{vtx}$ for signal candidates was required to lie in the~range $3000 < Z_{vtx} < 4700$ mm to avoid $\pi^0$'s generated by halo neutrons hitting detector components.
The~requirement on $P_t$ as a~function of $Z_{vtx}$ greatly suppresses the~background from $\mathrm{K_L}\to \pi^+ \pi^- \pi^0$ decays.
To reduce a~class of $\mathrm{K_L}\to \pi^0 \pi^0$ decays with miscombination of two photons in the~$\pi^0$ reconstruction, 
a requirement on the~photons energies ratio and the~product of their energy and the~angle between the~beam axis and their momenta is imposed.
The~opening angle cut in the~transverse plane was required to reduce the~$\mathrm{K_L}\to \gamma \gamma$ background, in which the~photons are back to back.
To select $\pi^0$ candidates with plausible kinematics, allowed regions were set on $(P_t/P_z,Z_{vtx})$ and $(E,Z_{vtx})$ planes, where $P_z$ and $E$ are the~longitudinal momentum and energy of the~$\pi^0$, respectively.
This cut was effective also to reduce the~background from $\eta$ mesons produced in the~halo-neutron interaction with the~of plastic scintillator veto-counter for charged particles (CV) \cite{Naito:2015vrz} located in front of CSI.
Other neutron-induced backgrounds are upstream $\pi^0$ and ``hadron cluster", caused by neutron directly hitting CSI and creating a~hadronic shower and by a~neutron produced in the~primary shower to create a~second, separated hadronic shower.
Table \ref{KOTOBkg} gives a~summary of the~different background contributions.

\begin{table}[h]
\caption{Summary of background estimation.}
 \begin{center}
   \begin{tabular}{l | c}
\hline
      source                 &  Expected events in signal region  \\
   \hline
   \multicolumn{2}{l}{$\mathrm{K_L}$ decay} \\  
   \hline
   $\mathrm{K_L}\to\pi^+\pi^-\pi^0$        & $0.05 \pm 0.02$ \\   
   $\mathrm{K_L}\to\pi^0\pi^0$             & $0.02 \pm 0.02$ \\
   Other $\mathrm{K_L}$ decays             & $0.03 \pm 0.01$ \\
   \hline
   \multicolumn{2}{l}{Neutron induced} \\
   \hline
   Hadron cluster                          & $0.24 \pm 0.17$ \\
   Upstream-$\pi^0$                        & $0.04 \pm 0.03$ \\
   CV-$\eta$                               & $0.04 \pm 0.02$ \\
   \hline \hline
   Total                                   & $0.42 \pm 0.18$ \\
   \hline
  \end{tabular}
 \end{center}
 \label{KOTOBkg}
\end{table}

\begin{figure}[h]
 \includegraphics[width=.5\textwidth]{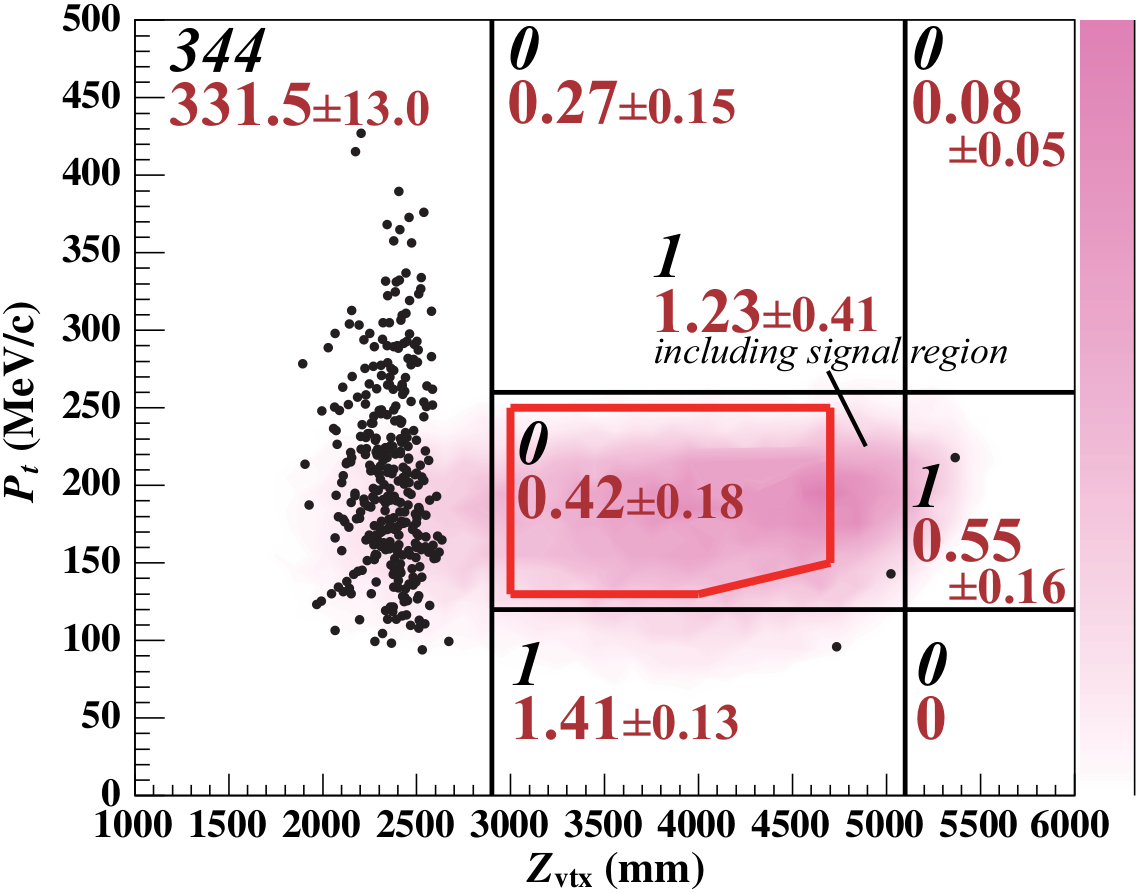}%
 \caption{Reconstructed  $\pi^0$ transverse momentum ($P_t$) as a~function of its decay vertex position $Z_{vtx}$.
          The~region surrounded by red lines is the~signal region. 
          Black markers represent observed events, and the~contour indicates the~$\mathrm{K_L}\to \pi^0 \nu \bar{\nu}$ signal distribution derived from the~MC simulation.
          The~black (red) numbers represent observed (expected background) events for the~regions inside the~lines. \label{KOTOFinal}}
\end{figure}

After full event selection no signal candidate events were observed, as shown in Fig. \ref{KOTOFinal}. 
Assuming Poisson statistics with uncertainties taken into account, the~upper limit on BR($\mathrm{K_L}\to \pi^0 \nu \bar{\nu}) < 3.0 \times 10^{-9}$ at 90\% CL was obtained \cite{Ahn:2018mvc},
improving the~upper limit of the~direct search by almost an order of magnitude.

The~data collected in years 2016-2018 are under study and the~results are expected soon, in the~summer of 2019.
The~preliminary estimates look very promising, with $SES = 8.2 \times 10^{-10}$ and total background of $0.20 \pm 0.16$.

\section{Other Kaon Decay Studies at NA62}

A~search for LFV/LNV in three track kaon decays was performed on data taken during three months in 2017, which corresponds to about 30\% of the~data collected in 2016-2018.
The~forbidden decays $\mathrm{K^+}\to\pi^-\mu^+\mu^+$ and $\mathrm{K^+}\to\pi^-e^+e^+$ were analyzed.
All the~details of the~analysis can be found in \cite{LFV}.
The~upper limits at 90\% CL are BR  $(\mathrm{K}^+ \to \pi^- e^+ e^+) < 2.2 \times 10^{-10}$ and BR$(\mathrm{K}^+\to \pi^- \mu^+ \mu^+) < 4.2 \times 10^{-11}$.
They are improving the~current PDG limits \cite{PDG2018} by a~factor of three in electron mode and by a~factor of two in muon mode.

\begin{figure}[h]
 \includegraphics[width=.5\textwidth]{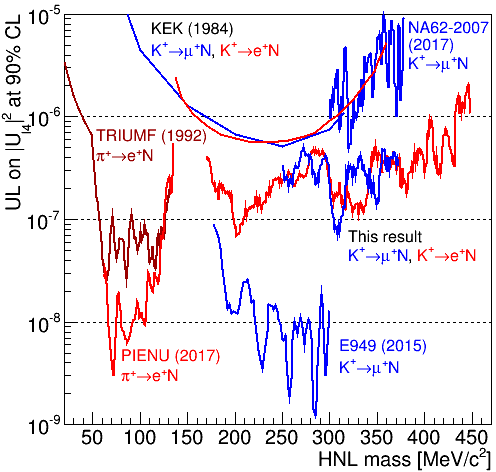}%
 \caption{ Upper limits on $|U_{\ell_4}|^2$ at~90\% CL obtained for each assumed HNL mass compared to other experimental results.\label{HNL}}
\end{figure}

A~search for HNL was completed on the~2015 data set and is published in \cite{CortinaGil:2017mqf}.
Upper limits on the~HNL mixing parameters $|U_{\ell_4}|^2$ at~90\% CL were established in the~ranges 170-448~MeV/$c^2$ and 250-373~MeV/$c^2$ for positron and muon modes, respectively, at a~level between $10^{-7}$ and $10^{-6}$, see Fig. \ref{HNL}.
Major improvements are foreseen with the~new data taken in years 2016-2018.

\section{Conclusion}
The~latest results on flavor physics golden modes $\mathrm{K}^+\to \pi^+ \nu \bar{\nu}$ and $\mathrm{K_L}\to \pi^0 \nu \bar{\nu}$ are reported.

The~charged mode is measured by the~NA62 experiment to set the~upper limits BR$(\mathrm{K^+}\to\pi^+\nu\bar{\nu}) < 14 \times 10^{-10}\ \mathrm{at}\ 95\%\ \mathrm{CL}$ \cite{CortinaGil:2018fkc}.
A~major improvement is foreseen when the~already collected data will be fully analyzed.

The~KOTO experiment measured the~neutral mode on 2013 and 2015 data sets.
The~upper limit on BR$(\mathrm{K_L}\to\pi^0\nu\bar{\nu})$ was obtained to be $<5.1\times 10^{-8}$ at 90\% CL based on 2013 data \cite{Ahn:2016kja}, and
BR($\mathrm{K_L}\to \pi^0 \nu \bar{\nu}) < 3.0 \times 10^{-9}$ at 90\% CL from 2015 data set \cite{Ahn:2018mvc}.

The~results on HNL and LFV/LNV studies by the~NA62 experiment are also discussed.
For LFV/LNV modes, the~upper limits at 90\% CL are BR  $(\mathrm{K}^+ \to \pi^- e^+ e^+) < 2.2 \times 10^{-10}$ and BR$(\mathrm{K}^+\to \pi^- \mu^+ \mu^+) < 4.2 \times 10^{-11}$ \cite{LFV}.
For HNL upper limits are set on the~HNL mixing parameters $|U_{\ell_4}|^2$ at~90\% CL \cite{CortinaGil:2017mqf}.

\begin{acknowledgments}
Support for~this work has been received from~the~grant LTT~17018
of~the~Ministry of~Education, Youth and~Sports of~the~Czech Republic
and from Charles University Research Center UNCE/SCI/013.
\end{acknowledgments}

\bigskip 

\end{document}